\documentclass[%
 twocolumn,
 groupedaddress,
 aps,
 pra,
]{revtex4-2}

\usepackage{graphicx}
\usepackage{dcolumn}
\usepackage{bm}
\usepackage{hyperref}
\usepackage{latexsym}
\usepackage{amsmath}
\usepackage{amsfonts}
\usepackage{amssymb}
\usepackage{amscd}
\usepackage{bbm}
\usepackage{bbold}
\usepackage{tikz}
\usepackage{tikz-cd}
\usetikzlibrary{shapes}
\usepackage{hyperref}
\usepackage[export]{adjustbox}
\usepackage[caption=false]{subfig}
\usepackage{soul}
\usepackage{comment}

\newcommand{\bra}[1]{\langle #1|}
\newcommand{\ket}[1]{|#1\rangle}
\newcommand{\braket}[2]{\langle #1|#2\rangle}
\newcommand{\ketbra}[2]{\vert #1 \rangle \! \langle #2 \vert}

\newcommand{\be}{\begin{equation}}
\newcommand{\ee}{\end{equation}}
\newcommand{\bea}{\begin{eqnarray}}
\newcommand{\eea}{\end{eqnarray}}

\newtheorem{defn}{Definition}

\newtheorem{lemma}{Lemma}
\newtheorem{prop}{Proposition}

\begin{document}


\title{A note on typicality in random quantum scattering}

\author{Michele Avalle}
\affiliation{Department of Computer Science, Sapienza University of Rome, Viale Regina Elena 295, Rome, 00161, Italy}

\author{Alessio Serafini}
\email[]{serale@theory.phys.ucl.ac.uk}
\affiliation{Department of Physics \& Astronomy, University College London, Gower Street, London WC1E 6BT, United Kingdom}

\date{\today}

\begin{abstract}
We consider scattering processes where a quantum system is comprised of an inner subsystem and of a boundary, and is subject to Haar-averaged random unitaries acting on the boundary-environment Hilbert space only. We show that, regardless of the initial state, a single scattering event will disentangle the unconditional state (i.e., the scattered state when no information about the applied unitary is available) across the inner subsystem-boundary partition. Also, we apply Levy's lemma to constrain the trace norm fluctuations around the unconditional state. Finally, we derive analytical formulae for the mean scattered purity for initial globally pure states, and provide one with numerical evidence of the reduction of fluctuations around such mean values with increasing environmental dimension.
\end{abstract}

\maketitle

\section{Typical states under restricted interactions}

Typicality arguments, based on Haar averages in the high-dimensional limit, have been repeatedly advocated as a plausible avenue to justify the second law and the ubiquity of the thermal state \cite{lubkin,gemmer01,gemmer09,popescuNature}. 
Such arguments rely on averages over Haar-distributed unitaries over a composite Hilbert space $\mathcal{H}_{\sigma \epsilon}=\mathcal{H}_{\sigma}\otimes\mathcal{H}_{\epsilon}$ 
pertaining to a physical system $\sigma$ interacting with an environment $\epsilon$. 
The typicality approach, closely related to the general study of typical partial entropies \cite{dahlsten14}, does allow one to shed light on the general thermodynamic behaviour of quantum systems without getting bogged down in unnecessary detail, and has been more recently extended to other quantifiers as well, such as quantum coherence \cite{singh16}. 
Besides, typical entanglement and states play an important role in quantum information science, both directly and through the construction of related $t$-designs, with applications to quantum cryptography, teleportation thresholds, state estimation, channel capacities and randomized benchmarking \cite{hayashi,iblisdir,hayden08,hayden09,emerson05,knill08,magesan12}.

However, it must be noted that actual 
physical interactions hardly ever comply with the notion of typicality set out above, 
as they do not extend to the whole system-plus-environment Hilbert space.
It is therefore worthwhile to reconsider the study of typical states and entropies by generalising the system/environment framework to one where the system is comprised of an inner part with Hilbert space $\mathcal{H}_I$ and of a boundary with Hilbert $\mathcal{H}_B$, and where only the latter interacts with the environment, with Hilbert space $\mathcal{H}_E$, through 
a Haar-distributed unitary over the space 
$\mathcal{H}_B\otimes\mathcal{H}_E$, as per the sketch below.
\begin{equation}
(\stackrel{I}{\bullet}\quad \underbrace{\stackrel{B}{\bullet})\quad \stackrel{E}{\bullet}}_{U} \nonumber \; .
\end{equation}
Somewhat loosely, we refer to this framework of restricted, `partial' mixing as `random scattering'. This framework will allow us to inquire into how equilibration \cite{popescupre} and thermalisation play out for many-body systems with only a partial interface interacting with the environment.


Notably, a set-up similar to the above one plays a central role in the seminal Hayden-Preskill model for information retrieval from a black hole's Hawking radiation. Therein, the internal dynamics of a black hole is modelled through a random, Haar-distributed unitary acting only on two specific subsystems (which are maximally entangled with two other reference systems). Building on previous work \cite{hayden08,hayden09}, Hayden and Preskill make use of an upper bound constraining the trace distance between the averaged system reference state and a product state, given in terms of dimensions and inner state purity. Such structured black hole models, based on the selective application of `pseudorandom' interactions, are currently drawing substantial attention \cite{yoshida19,piroli20,akers22,preskill23}.
It is also worthwhile to mention that, in a similar vein, pseudorandom interactions are finding application in other fundamental investigations, such as the study of wormhole growth in the AdS/CFT correspondence \cite{bouland19}.

In this work, we determine the average state resulting from a scattering event, and show that it is always completely uncorrelated across the inner subsystem-boundary partition; also, by applying a standard argument based on Levy's lemma, we exponentially bound the fluctuations around such a mean state for initial states that are separable across the inner subsystem-boundary partition, and for any initial state of the environment. 
Next, we determine the average purity of the system (inner plus boundary) state under Haar-averaged scattering interactions for an initial global pure state, as well as for an initial factorised state of the environment as a function of the initial inner subsystem-boundary Schmidt coefficients and subsystems' dimensions alone. 

The plan of the paper is as follows:
in Section \ref{sec:map} we define the random scattering process, set out the basic terminology and introduce some of the quantities adopted; in Section \ref{sec:unco} we determine the average unconditional state and study its fluctuations for initially separable states; 
in Section \ref{sec:purity} we determine the mean local purity of initially pure scattered states and, by addressing numerically GHZ and W states, present evidence that the fluctuations around such mean values shrink with increasing environmental dimension;
in Section \ref{sec:conclu} we draw some conclusions; derivations and preliminary technical material concerning Haar averages, Levy's lemma and concentration of measure are deferred to the appendices.

\section{The scattering map}\label{sec:map}

Let us consider the scattering map $\Phi$ sending the global initial state $\rho$, defined on the Hilbert space $\mathcal{H}_I\otimes\mathcal{H}_B\otimes\mathcal{H}_E$, into the state of subsystem $IB$ after a random scattering interaction acting on subsystem $BE$ alone:
\begin{equation}\label{eq:map}
\Phi(\rho) = \hbox{Tr}_{E}\left[\int_{H}\hbox{dU}_{BE}\left(\mathbb{I}_I\otimes U_{BE}\right)\rho\left(\mathbb{I}_I\otimes U_{BE}\right)^{\dagger}\right] \, ,
\end{equation}
where the Haar average over the unitary group, denoted with 
$\int_H \hbox{dU}_{BE}$, takes into account the lack of information about the boundary-environment interaction in a random scattering process. 
To emphasise the Haar averaging involved, we shall also refer to the scattered 
system state as $\langle\rho_{IB}\rangle=\Phi(\rho)$ in what follows. Basic techniques for the evaluation of Haar averages are sketched in Appendix \ref{app:measure}.

As customary in this context, we shall adopt the purity $\mu=\hbox{Tr}(\rho^2)$ as an entropic measure. The purity is related to the so-called linear entropy $S_L=(1-\mu)$,
which can in turn be derived from the von Neumann entropy $S_V=-\hbox{Tr}(\rho\ln \rho)$ through a Taylor expansion at lowest order in the quantum state.
At variance with the von Neumann entropy, the purity is not endowed with a direct operational interpretation but is much more expedient to evaluate and is still very effective in characterising the mixedness and thermal character of a state, especially around its extremal values ($1$ for pure states and $1/d$ for maximally mixed states, $d$ being the system dimension). 
It is easy to see directly from its definition that the purity is a convex function on the set of quantum states.

In the following we will consider both the 
purity $\hbox{Tr}\left(\Phi(\rho)^2\right)$ 
of the scattered state, as well as the average of the purity 
under the Haar measure
\begin{equation}\label{eq:puritydef}
\langle\mu(\rho_{IB})\rangle =\hbox{Tr}_{IB}\left[ \int_{H}\hspace*{-0.18cm}\hbox{dU}\left(\hbox{Tr}_{E}\left(\mathbb{I}_I\otimes U\right)\rho\left(\mathbb{I}_I\otimes U\right)^{\dagger}\right)^2 \right] , 
\end{equation}

Both $\langle \rho_{IB}\rangle$ and $\langle \mu(\rho_{IB})\rangle$ can be evaluated exactly and, as we will see in the next sections,
the fluctuations around them can be bound by means of Levy's lemma.

Borrowing from well-established terminology in quantum control theory, we will refer henceforth to the state $\langle \rho_{IB}\rangle$ as to the `unconditional' state, which is prepared 
after a scattering event when the interaction between boundary and environment is completely unknown, and thus Haar-distributed. The unconditional state is an average of conditional 
states $\rho_{IB}$, which would be prepared if information about the unitary interaction were somehow retrieved. The purity distribution of conditional states may be of interest too, and will therefore be considered in what follows.

\section{The unconditional state}\label{sec:unco}

Let us first analyse some remarkable properties of the unconditional state, resulting from the Haar average, and consider their thermodynamical implications. 
Throughout the paper, we will use a convention
such that the indices of all matrices and coefficients will always follow the order:
inner system-boundary-environment. Also, we shall set 
$d_X:=\hbox{dim}(\mathcal{H}_X)$.

\subsection{A decoupling theorem} \label{sec:map}

Quite remarkably, the scattering map of Eq.(\ref{eq:map}) has the effect of totally suppressing any initial correlation between
the boundary and the inner part of the system.
\begin{prop} \label{prop:map}
For any given initial state $\rho$, the expectation value over the Haar measure of the reduced $IB$
state after the scattering interaction defined in Eq.(\ref{eq:map}) is a factorised state:
\begin{equation} \label{eq:propmap}
\langle\rho_{IB}\rangle=\rho_I\otimes \frac{\mathbb{I}_B}{d_B}\, ,
\end{equation}
with 
\begin{equation} \label{eq:propmap2}
 \rho_I=\hbox{Tr}_{BE}(\rho) \; .
\end{equation}
\end{prop}
The proof of this proposition is deferred to Appendix \ref{app:decoupling}.

Notice that the local parts of this unconditional state might have been predicted by inspection, since the maximally mixed state in the boundary is a result of the Haar-averaged mixing with the environment, whilst the inner system is untouched by the interaction, which thus will not alter its local state. However, the complete destruction of correlations, regardless of the dimensions involved, is not trivial. 
This proposition complies with what is known from the study of Hawking radiation \cite{haydenpreskill}, in a form and to a degree that will be made more explicit in the next subsection. 

Before moving on to quantifying the statistical deviation from the unconditional state, it is worthwhile to spend a few words about its thermodynamic implications. 
If one takes the stance that thermalisation can be justified by localised, yet unknown, surface interactions with an environment, this statement shows that environmental thermalisation must be entirely mediated by the surface, i.e.~it cannot act on the bulk of a system directly through previous correlations, regardless of the subsystem sizes at play.

\subsection{Fluctuations around the unconditional state}

Levy's lemma (see Appendix \ref{app:levy} for a discussion of its derivation) is a powerful standard tool to 
characterise the Haar-generated distribution of conditional states. 
Let us first recall the lemma:
\begin{lemma} \label{lemma:levy}
(\textbf{Levy's lemma}) Given a function $f$ : $S^d\rightarrow \mathbbm{R}$ defined on the d-dimensional hypersphere $\mathbbm{S}^d$,
and a point $\phi \in \mathbbm{S}^d$ chosen at random,
the probability $P$ for $f$ to deviate from its mean value is given by
\begin{equation} \label{eq:levydef}
\hbox{P}\left[|f(\phi)-\langle f\rangle|\geq \epsilon\right]\leq 2\hbox{exp}\left(-\frac{(d+1)\epsilon^2}{9\pi^3\eta^2} \right) \, ,
\end{equation}
where $\epsilon$ is an arbitrarily small positive constant and $\eta$ is the Lipshitz constant of $f$, i.e. 
$\eta$ : $|f(\phi_1)-f(\phi_2)|\leq \eta |\phi_1-\phi_2|$, $\forall (\phi_1,\phi_2)\in \mathbbm{S}^d$.
\end{lemma}
The above lemma can be applied anytime one deals with pure global quantum states which, for a Hilbert space of dimension $d$, live on the surface of a $(2d-1)$ dimensional hypersphere, and is in fact key to thermodynamic typicality arguments, as in \cite{popescuNature}.  
In our case though, the action of the random, Haar-averaged unitaries is constrained to the $IB$ subspace, so we 
cannot apply the lemma directly.

Yet, let us consider an initial state that is separable across the inner-boundary sector, i.e., a state of the form
\begin{align}
\rho =& \sum_{j} p_j \rho_{I,j}\otimes \rho_{BE,j} \\
=& 
\sum_{j,k} p_j p_{BE,j,k} 
\rho_{I,j}\otimes \ket{\psi_{BE,j,k}}\bra{\psi_{BE,j,k}} \, ,
\end{align}
where $p_j$ and $p_{BE,j,k}$ are positive probabilities adding up to one when summed over $j$ and $k$ respectively, the $\rho_{I,j}$'s are quantum states of the inner sector and $\ket{\psi_{BE,j,k}}$ are unit vectors of the Hilbert space
$\mathcal{H}_B\otimes\mathcal{H}_E$.
The action of the Haar-averaged unitaries on each element the sum above  
results in a probability distribution of pure states
in $\mathcal{H}_B\otimes\mathcal{H}_E$ that does not depend on $j$ and $k$, and which may be  parametrised on the $(2d_Bd_E-1)$-dimensional hypersphere of pure states on such a space.
Denoting by $\phi$ the variables that parametrise the hypersphere, and by $\ket{\phi}$ the corresponding pure state of the $BE$ subsystem, one has that the initial state $\rho$ is mapped into the following conditional state 
\be\label{conditional}
\sum_{j,k} p_j p_{BE,j,k} 
\rho_{I,j}\otimes \ket{\phi}\bra{\phi} = 
\rho_I\otimes\ket{\phi}\bra{\phi}\, ,
\ee
where $\rho_I= \sum_{j}p_j\rho_{I,j} =\hbox{Tr}_{BE}(\rho)$.
Then, 
as proven in detail in Appendix \ref{app:flucs}, Levy's lemma may be applied to obtain the following characterisation of the distribution of trace distances of the scattered states
\begin{equation} \label{eq:lastlevy}
P\left[ \|\rho_{IB,\phi}-\Phi \left( \rho\right) \|_1 \geq \epsilon + \sqrt{\frac{d_B^2-1}{d_Ed_B+1}}\right] \leq 2\hbox{e}^{-\frac{d_Bd_E\epsilon^2}{18\pi^3}} \, . 
\end{equation}
Let us also recall that, by virtue of Helstrom's theorem \cite{helstrom}, the minimum error probability in discriminating between two quantum states $\rho$ and $\sigma$, optimised over all POVMs, is given by $\frac12-\frac14\|\rho-\sigma\|_1$. In this precise sense, states at vanishing trace distance become operationally indistiguishable, an argument that was also applied to justify perceived thermodynamic regularities in the appropriate limit 
\cite{popescuNature}.

Inequality (\ref{eq:lastlevy}) only applies to separable inner-boundary initial states (which also include cases where the environment is initially completely uncorrelated, although this requirement was not explicitly needed), as Levy's lemma does not allow one to make any direct inference for initial entangled states, where the effect of random unitaries on off-diagonal elements must be taken into account. Under such a separability assumption, the dimension of the inner Hilbert space does not play any role in bounding the typical fluctuations. We will explore more general initial states and address situations where the inner Hilbert space does play a role, by investigating the purity distribution of the scattered states in the next section.

Notice also that, as typical in such cases \cite{popescuNature}, our argument does not demonstrate an exponential shrinking of the deviation from the average when the trace distance is arbitrarily small, although this is recovered in the limit $d_E\gg d_B$: As one should expect, concentration of measure occurs, even under partial averaging, as the environment dimension grows, regardless of the other dimensions involved.

\section{Mean scattered purity of initial pure states} \label{sec:purity}

A scenario which is amenable to an insightful evaluation is that of initial pure states. Denoting with $\{\ket{i}\}$, $\{\ket{b}\}$ and $\{\ket{e}\}$ the bases of choice in the spaces $\mathcal{H}_I$, $\mathcal{H}_B$ and $\mathcal{H}_E$, and with $\psi_{ibe}$ arbitrary state vector coefficients, one would have for pure states: 
\begin{equation} \label{eq:initrho}
\rho=\ketbra{\Psi}{\Psi}=\sum_{ii'bb'ee'}\psi_{ibe}\psi_{i'b'e'}^*\ketbra{ibe}{i'b'e'} \, .
\end{equation}
Through a rather lengthy calculation, reported in Appendix (\ref{sec:puritycal}),
one then obtains the following mean purity
\begin{equation}\label{eq:purityres}
\langle \mu(\rho_{IB})\rangle = \frac{d_B+d_E}{d_Bd_E+1}+\frac{d_B(1-d_E^2)}{(d_Bd_E)^2-1}\left( \Delta - \Gamma\right)\, ,
\end{equation}
where we have defined
\begin{eqnarray}
 \Delta & := & \sum_{\substack{i,i',(be),(be)'\\ i\neq i',(be)\neq (be)'}}|\psi_{i(be)}|^2|\psi_{i'(be)'}|^2\, , \\
 \Gamma & := & \sum_{\substack{i,i',(be),(be)'\\ i\neq i',(be)\neq (be)'}}\psi_{i(be)}\psi_{i'(be)}^*\psi_{i'(be)'}\psi_{i(be)'}^* \, .
\end{eqnarray}

When the initial system state is pure, and the system-environment state is factorised, the only additional parameters at play, other than the Hilbert spaces' dimensions, are the Schmidt coefficients of the $IB$ initial state, on which the mean purity must necessarily depend. As shown in Appendix \ref{app:schmidt}, the latter then reads 
\begin{equation}\label{eq:schmidt}
\langle \mu(\rho_{IB})\rangle=\frac{d_B+d_E}{d_Bd_E+1}+\frac{d_B(1-d_E^2)}{(d_Bd_E)^2-1}\sum_{\substack{jl\\ j\neq l}}|c_j|^2|c_l|^2 \, ,
\end{equation} 
where $\{c_j\}$ are the Schmidt coefficients of the initial $IB$ state.

Clearly, Eq.~(\ref{eq:schmidt}) implies  that, when the initial state of the $IB$ system is also separable,
i.e., when $\ket{\Psi}=\ket{I}\otimes\ket{B}\otimes\ket{E}$, $M=1$, one recovers the well-known result in the absence of inner system-boundary separation \cite{muller,dahlsten14}:
\begin{equation}\label{eq:trivialpurity}
\langle \mu(\rho_{IB})\rangle=\frac{d_B+d_E}{d_Bd_E+1}\, . 
\end{equation} 
This is actually true regardless the initial correlations between boundary and environment. Indeed, when 
$\ket{\Psi}=\ket{I}\otimes \ket{BE}=\sum_{i=1}^{d_I}\sum_{j=1}^{d_Bd_E}\gamma_i\tau_j\ket{\Gamma_i}\ket{T_j}$:
\begin{eqnarray}
\Delta - \Gamma & = & \sum_{ii'jj'}\left[ |\gamma_i\tau_j|^2|\gamma_{i'}\tau_{j'}|^2-\gamma_i\tau_j\gamma_{i'}^*\tau_j^*\gamma_{i'}\tau_{j'}\gamma_i^*\tau_{j'}^*\right] \nonumber \\
& = & \sum_{ii'jj'}\left[ |\gamma_i\tau_j|^2|\gamma_{i'}\tau_{j'}|^2-|\gamma_i\tau_j|^2|\gamma_{i'}\tau_{j'}|^2\right] \nonumber \\
& = & 0 \, ,
\end{eqnarray}
so that Eq.~(\ref{eq:purityres}) reduces to the standard bipartite system-environment case.

\begin{figure*}[ht!]
\makebox[\textwidth][c]{
{\includegraphics[scale=0.5]{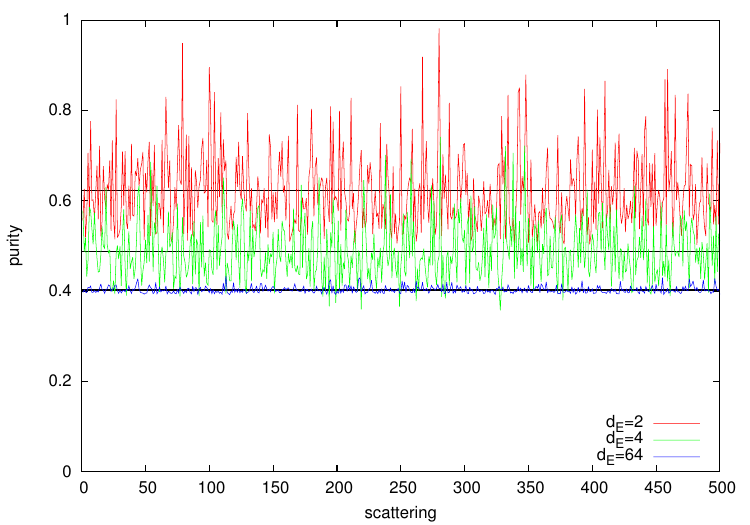}}
{\includegraphics[scale=0.5]{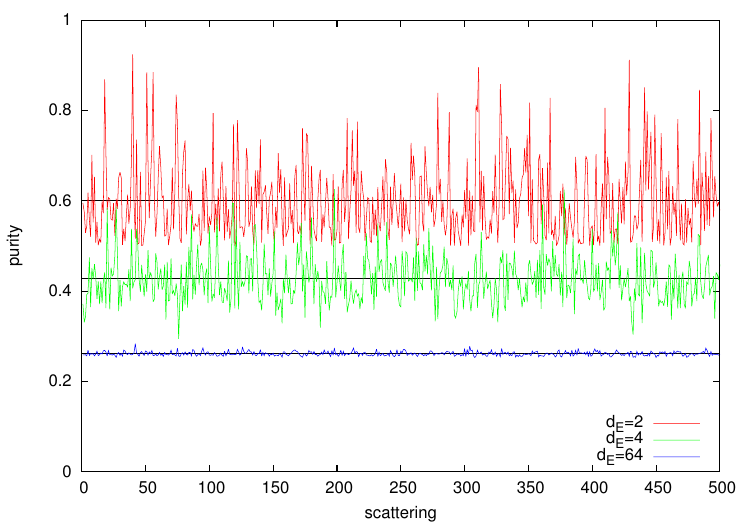}}}

\makebox[\textwidth][c]{
{\includegraphics[scale=0.5]{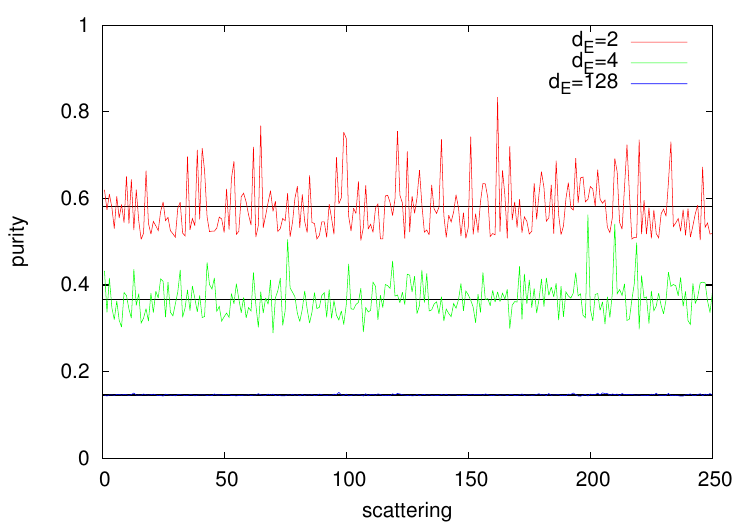}}
{\includegraphics[scale=0.5]{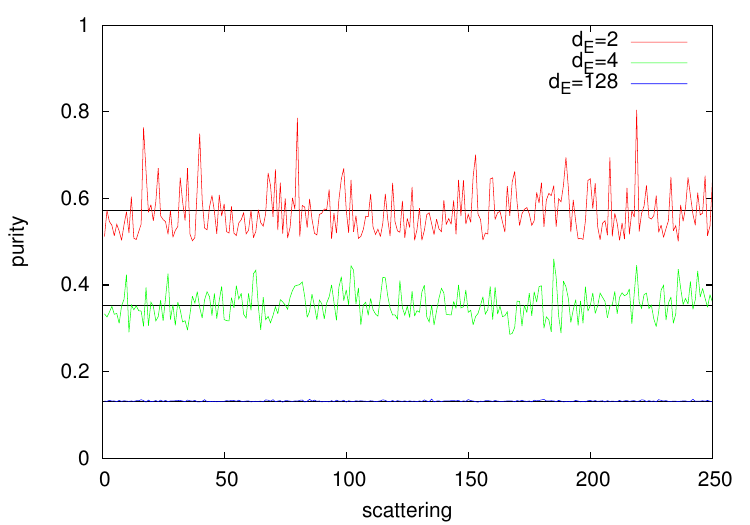}}}
\caption{Purity of the state resulting from a random scattering boundary-environment interaction for (see Eqs.(\ref{eq:WGHZ},\ref{eq:WGHZE})) (a) $\rho^{W}$,
(b) $\rho^{GHZ}$, (c) $\tilde{\rho}^{W}$, (d) $\tilde{\rho}^{GHZ}$ initial states.
Each value on the $x$ axis labels a different $U_{BE}$ random extraction.
The straight horizontal lines in the plots represent the expected values of the purity, computed with Eqs.(\ref{eq:WGHZ}),
\ref{eq:WGHZE}, where in (a),(b) $d_I=d_B=2$, whereas in (c),(d) $d_I=2$, $d_B=4$.}
\label{fig:Wghz}
\end{figure*}

If, on the other hand, the initial inner system-boundary state is maximally entangled, assuming, as is reasonable, $d_B\le d_I$, one can insert $c_j=1/\sqrt{d_B}$ $\forall$ $j$ to obtain, from Eq.~(\ref{eq:schmidt}), the mean purity
\be
\langle \mu(\rho_{IB})\rangle = 
\frac{(d_B^2-1)d_E+d_E^2-1}{(d_B^2d_E^2-1)} \; ,
\ee
which, in the limit $d_E\rightarrow\infty$, yields 
$\langle \mu(\rho_{IB})\rangle = 1/d_B^2$: the boundary's dimension 
constrains the equilibrium mean purity 
to a value higher than the minimum  value $1/(d_Id_B)$.

This corresponds to the fact that full mixing (and, in settings where the total energy is set, full thermalisation) 
cannot be achieved via interaction with a boundary, unless further boundary-inner subsystem interactions are also taken into account.

The fundamental difference between our scattering model and the conventional scenario in which the whole system interacts with an environment
lies in the fact that in the latter case the expectation value of the purity is independent of the initial state, whilst in the former case, in general, this
is clearly not true. This is the case even in the limit of infinite dimension of the environment:
\begin{equation}
\lim_{d_E \to \infty}\langle \mu(\rho_{IB})\rangle=\frac{1}{d_B}\left(1-\Delta+\Gamma\right)\, .
\end{equation}
\\

Here, Levy's lemma cannot be applied directly to bound the fluctuations around these mean values, as the random unitary transformations do not in general result in a state distribution that can be parametrised on a hypersphere.
However, 
we have checked numerically the validity of Eqs.~(\ref{eq:purityres}) foro initial GHZ and W states. For $N$ qubits, these two classes of states, which are important for the study of multipartite entanglement \cite{horodecki},
are defined as:
\begin{eqnarray}
 \ket{GHZ}_N & = & \frac{1}{\sqrt{2}}\left(\ket{0}^{\otimes N}+\ket{1}^{\otimes N}\right)\, , \\
 \ket{W}_N & = & \frac{1}{\sqrt{N}}\left(\ket{10\dots 0}+\ket{01\dots 0}+\dots+\ket{00\dots 1} \right). \nonumber \\
\end{eqnarray}
Defining $\rho^{GHZ}=\ketbra{GHZ}{GHZ}_N$ and $\rho^{W}=\ketbra{W}{W}_N$, a direct calculation leads to the following mean scattered purities: 

\begin{eqnarray} \label{eq:WGHZ}
\langle \mu(\rho_{IB}^{GHZ})\rangle & = & \frac{d_B+d_E}{d_Bd_E+1}+\frac{d_B(1-d_E^2)}{(d_Bd_E)^2-1} \frac{1}{2}   \nonumber \\
\langle \mu(\rho_{IB}^{W})\rangle & = & \frac{d_B+d_E}{d_Bd_E+1}+\frac{d_B(1-d_E^2)}{(d_Bd_E)^2-1} \frac{2N_I(N-N_I)}{N^2} \; , \nonumber\\
\end{eqnarray}
where $D=d_Id_Bd_E=2^N$ (here, the $X$ subsystem comprises $N_X=\log_2(d_X)$ qubits). Interestingly, the average scattered purity of W states is affected by the inner subsystem dimension. Also, it can be seen that the average purity of initial W states is larger than the GHZ one (for given $d_B$ and $d_E$) as long as $N_I<N/2$, i.e., as long as the boundary interaction Hilbert space is not too thin compared with the system bulk.

We have also considered the case where three-qubit GHZ and W states interact with an $N_E=N-3$ qubit environment prepared in a fiducial state
$\tilde{\rho}^{GHZ}:=\ketbra{GHZ}{GHZ}_3\otimes \ketbra{0}{0}_E^{\otimes (N-3)}$,
$\tilde{\rho}^{W}:=\ketbra{W}{W}_3\otimes \ketbra{0}{0}_E^{\otimes (N-3)}$, finding
\begin{eqnarray} \label{eq:WGHZE}
\langle \mu(\tilde{\rho}_{IB}^{GHZ})\rangle&=&\langle \mu(\rho_{IB}^{GHZ})\rangle \, , \\ \langle \mu(\tilde{\rho}_{IB}^{W})\rangle &=& \frac{d_B+d_E}{d_Bd_E+1}+\frac{d_B(1-d_E^2)}{(d_Bd_E)^2-1} \frac{4}{9} \, .
\end{eqnarray}
Note that these formulae apply to both $N_I=1$, $N_B=2$ and to $N_I=2$, $N_B=1$, since the Schmidt coefficients across the inner boundary partition are the same in both set-ups.

As can be appreciated in Fig.(\ref{fig:Wghz}), the fluctuations around the expected value of $\mu$ are quite large
for an environment composed of $N_E=1$, $N_E=2$ qubits, but are already strongly suppressed for $N_E\approx 10$ and almost completely damped for
$N_E\approx 100$.

\section{Conclusions}\label{sec:conclu}
Summing up, we have considered a scattering scenario where random unitary transformations act jointly on an environment and a boundary subsystem rather than on the whole system, which comprises an inner subsystem too, and derived the following:
\begin{itemize}
\item A proof that any single, Haar-distributed scattering event is able to disentangle completely boundary and inner subsystems.
\item A proof that, for initially separable inner-boundary states, the conditional 
scattered states concentrate exponentially in trace norm around the average conditional state, by virtue of Levy's lemma.
\item An analytical formula for the mean purity of initially pure scattered states in terms of the subsystems' dimensions and initial state's coefficients.
\item An analytical formula for the mean scattered purity of initially globally pure states uncorrelated with the environment, in terms of the Schmidt coefficients of the initial state. 
\end{itemize}
Besides, by addressing initial GHZ and W states, we have provided numerical evidence for the concentration of measure of scattered purities with increasing environmental dimension.

Our approach and findings are a step towards a generalised study of thermalisation and equilibration for structured systems and restricted interactions, and may be of interest to quantum thermodynamics approaches based on typicality and other fundamental investigations, such as the study of black hole entropy.

\begin{acknowledgments}
We wish to thank the Madonna del Carmelo for xer unrelenting support.
\end{acknowledgments}

\widetext
\appendix

\section{Integrals over the unitary group} \label{app:measure}

Here we just summarise the results of \cite{aubert} which are relevant for the purposes of our model. 
\\

The maps Eqs.(\ref{eq:map},\ref{eq:puritydef}) involve the calculation of integrals of the form
\begin{equation}
 \int_H\hbox{dU}\left[ U^*_{i,j}U_{k,l}\right] \, , \quad \int_{H}\hbox{dU}\left[U^*_{i_1,j_1}U^*_{i_2,j_2}U_{k_1,l_1}U_{k_2,l_2} \right]\, ,
\end{equation}
where $\int_H$ means integrating over the Haar measure.
Integrals of this kind can be tackled using Schur's lemma (see e.g. \cite{schur,muller}), but here we follow the, somewhat easier, approach described in \cite{aubert}.

In general, for some degrees of the polynomials $p$, $q$ in $U^*$ and $U$, one wants to compute
\begin{equation}
 \int_H\hbox{dU}\left[U^*_{i_1,j_1}\dots U^*_{i_p,j_p}U_{k_1,l_1}\dots U_{k_q,l_q}\right]  =  \int_H\hbox{dU}\left[ \prod_{a=1}^pU^*_{i_a,j_a}\prod_{b=1}^qU_{k_b,l_b}\right]
                    \equiv  \int_H\hbox{dU}\left[U^*_{I_pJ_p}U_{K_qL_q}\right]
                    \equiv  \braket{I_p,J_p}{K_q,L_q}\, ,
\end{equation}
where we have defined $X_p=(x_1,x_2,\dots x_p)$. It is shown in \cite{aubert} that the only non-zero integrals are the ones in which
the degrees are such that $p=q$ (thus we will drop this index), $K=I$ and $L=J_Q$, where $J_Q$ is any permutation of the indices in the set $J$:
\begin{equation}
 \braket{I_p,J_p}{K_q,L_q}=\braket{I,J}{I,J_Q}\, .
\end{equation}

When $p=1$ 
\begin{equation}
 \int_H\hbox{dU}\left[ U^*_{i,j}U_{k,l}\right]=\int_H\hbox{dU}\left[ U^*_{i,j}U_{i,j}\right]=\braket{i,j}{i,j}=\frac{1}{d}\, ,
\end{equation}
where $d$ is the dimension of $U$.
\\

When $p=2$, the non-zero integrals are:
\begin{eqnarray} 
 \bullet \, (i_1\neq i_2,j_1\neq j_2) & : & \int_H\hbox{dU}\left[U^*_{i_1,j_1}U^*_{i_2,j_2}U_{i_1,j_1}U_{i_2,j_2} \right]=\frac{1}{d^2-1}\, ; \label{eq:nonzero1} \\
 \bullet \, (i_1\neq i_2,j_1\neq j_2) & : & \int_H\hbox{dU}\left[U^*_{i_1,j_1}U^*_{i_2,j_2}U_{i_1,j_2}U_{i_2,j_1} \right]=-\frac{1}{d(d^2-1)}\, ; \label{eq:nonzero2} \\
 \bullet \, (i_1=i_2,j_1\neq j_2) & : & \int_H\hbox{dU}\left[U^*_{i_1,j_1}U^*_{i_1,j_2}U_{i_1,j_1}U_{i_1,j_2} \right]=\frac{1}{d(d+1)}\, , \nonumber \\ 
                                  &   & \int_H\hbox{dU}\left[U^*_{i_1,j_1}U^*_{i_1,j_2}U_{i_1,j_2}U_{i_1,j_1} \right]=\frac{1}{d(d+1)}\, ; \label{eq:nonzero3}\\
 \bullet \, (i_1\neq i_2,j_1=j_2) & : & \int_H\hbox{dU}\left[U^*_{i_1,j_1}U^*_{i_2,j_1}U_{i_1,j_1}U_{i_2,j_1} \right]=\frac{1}{d(d+1)}\, ; \label{eq:nonzero4}\\
 \bullet \, (i_1=i_2,j_1=j_2) & : & \int_H\hbox{dU}\left[U^*_{i_1,j_1}U^*_{i_1,j_1}U_{i_1,j_1}U_{i_1,j_1} \right]=\frac{2}{d(d+1)}\, . \label{eq:nonzero5}\\ \nonumber
\end{eqnarray}

Knowing how to deal with this kind of integrals allows one to compute relevant quantities such as the mean reduced state $\overline{\rho_S}$ of a 
bipartite system-environment ($SE$) density matrix:
\begin{small}
\begin{eqnarray}
 \overline{\rho_S} & = & \hbox{Tr}_E(\langle \rho_{SE}\rangle_H)
               =  \hbox{Tr}_E\left\{\int_H\hbox{dU}\left[U\rho_{SE}U^{\dagger}\right]\right\}\nonumber\\
               & = &  \hbox{Tr}_E\left\{\int_H\hbox{dU}\sum_{acge=1}^{d_S}\sum_{ik=1}^{d_S}\sum_{bdhf=1}^{d_E}\sum_{jl=1}^{d_E}U_{(ab)(cd)}U^*_{(ef)(gh)}\psi_{ij}\psi^*_{kl}
              \ketbra{ab}{cd}\cdot\ketbra{ij}{kl}\cdot\ketbra{gh}{ef}\right\}\nonumber\\
               &=&  \hbox{Tr}_E\left\{\sum_{ikae=1}^{d_S}\sum_{jlbf=1}^{d_E}\psi_{ij}\psi^*_{kl}\underbrace{\int_H\hbox{dU}\left[ U_{(ab)(ij)}U_{(ef)(kl)}^*\right]}_{=\frac{1}{d_Sd_E}\delta^{ab}_{ef}\delta^{ij}_{kl}}\right\}\nonumber\\
               & = & \hbox{Tr}_E\left\{\frac{1}{d_Sd_E}\sum_{ij}|\psi_{ij}|^2\sum_{a=1}^{d_S}\sum_{b=1}^{d_E}\ketbra{ab}{ab}\right\}
               =  \frac{1}{d_Sd_E}\sum_{\xi=1}^{d_E}\bra{\xi}\sum_{a=1}^{d_S}\sum_{b=1}^{d_E}\ketbra{ab}{ab}\cdot\ket{\xi}
               =  \frac{1}{d_S}\sum_{a=1}^{d_S}\ketbra{a}{a}\, .
\end{eqnarray}
\end{small}

\section{Proof of proposition \ref{prop:map}}\label{app:decoupling}

{\em Proof:}
Due to the linearity of averaging and partial tracing, it will suffice to prove 
this statement for initially pure states.
So, substituting Eq.(\ref{eq:initrho}) into Eq.(\ref{eq:map}) and expressing the scattering unitaries in terms of their components
in the $BE$ basis, $U_{BE}=\sum_{i,k=1}^{d_B}\sum_{j,l=1}^{d_E} U_{(ij)(kl)}\ketbra{ij}{kl}$, yields
\begin{eqnarray} \label{eq:mapresult}
\Phi(\rho_{IBE}) & = & \hbox{Tr}_{E}\left[\int_{H}\hbox{dU}\left(\mathbb{I}_I\otimes U_{BE}\right)\rho_{IBE}\left(\mathbb{I}_I\otimes U_{BE}\right)^{\dagger}\right] \nonumber \\
& = & \sum_{\xi =1}^{d_E}\bra{\xi} \int_{H}\hbox{dU} \sum_{\substack{jklm\\ mnop}}\sum_{\substack{sbe\\ s'b'e'}}U_{(ij)(kl)}\psi_{sbe}\ket{ij}\langle{kl}\ket{sbe}\langle{s'b'e'}
 \ketbra{mn}{op}\psi_{s'b'e'}^*U_{(op)(mn)}^*\ket{\xi} \nonumber \\
& = &\sum_{\xi =1}^{d_E} \sum_{\substack{jklm\\ mnop}}\sum_{\substack{sbe\\ s'b'e'}}\psi_{sbe}\psi_{s'b'e'}^*\underbrace{\bra{\xi}{ij}\rangle\langle{kl}\ketbra{sbe}{s'b'e'}{mn}\rangle\langle{op}\ket{\xi}}_{\delta_j^{\xi}\delta^k_b\delta_e^l\delta_{b'}^m\delta_{e'}^m\delta_o^{\xi}}  \int_{H}\hbox{dU}\left[ U_{(op)(mn)}^* U_{(ij)(kl)}\right] \nonumber \\
& = & \sum_{\xi=1}^{d_E}\sum_{\substack{sbe\\ s'b'e'}}\sum_{io}\psi_{sbe}\psi_{s'b'e'}^*\ketbra{si}{s'o}\int_{H}\hbox{dU}\left[ U_{(o\xi)(b'e')}^* U_{(i\xi)(be)}\right] 
 =  \sum_{\xi=1}^{d_E}\sum_{\substack{sbe\\ s'b'e'}}\sum_{io}\psi_{sbe}\psi_{s'b'e'}^*\ketbra{si}{s'o}\frac{1}{d_Bd_E}\delta_i^o\delta_{b'e'}^{be} \nonumber \\
& = & \frac{1}{d_Bd_E}\sum_{\xi ss'bei}\psi_{sbe}\psi_{s'be}^*\ketbra{si}{s'i} 
 =  \sum_{ss'be}\psi_{sbe}\psi_{s'be}^*\ketbra{s}{s'} \otimes \frac{1}{d_B}\sum_{i=1}^{d_B}\ketbra{i}{i} 
 =  \rho_I \otimes \overline{\rho}_B \, ,
\end{eqnarray}
where for the fourth equality we have used the results of Appendix \ref{app:measure}.
\begin{flushright}
 $\square$
\end{flushright}

\section{Expectation value of the purity} \label{sec:puritycal} 
Here we explicitely show how to get to Eq.(\ref{eq:purityres}).
Substituting Eq.(\ref{eq:initrho}) into Eq.(\ref{eq:puritydef}) and expressing the scattering unitaries in terms of their components
in the $BE$ basis, $U_{BE}=\sum_{i,k=1}^{d_B}\sum_{j,l=1}^{d_E} U_{(ij)(kl)}\ketbra{ij}{kl}$:
\begin{eqnarray}
 \langle \mu(\rho_{IB})\rangle & = & \hbox{Tr}_{IB}\left\{ \int_{H}\hbox{dU}\left[ \hbox{Tr}_E\left(\left(\mathbb{I}_I\otimes U_{BE}\right)\rho_{IBE}\left( \mathbb{I}_I\otimes U_{BE}\right)^{\dagger}\right)\right]^2\right\} \nonumber \\
 & = & \int_H\hbox{dU} \sum_{\gamma\gamma'\beta\beta'\xi\xi'}\bra{(\gamma\beta)\xi} \sum_{\substack{ibe \\ i'b'e'}}
 \sum_{\substack{jklm \\ nopq}}\psi_{ibe}\psi_{i'b'e'}^*U_{(jk)(lm)}U^*_{(pq)(no)}\ket{jk}\langle{lm} 
\ket{ibe}\langle{i'b'e'}\ket{no}\langle{pq}\ketbra{(\gamma'\beta')\xi}{(\gamma'\beta')\xi'}\nonumber\\
 &\times&\sum_{\substack{i^2b^2e^2 \\ i^3b^3e^3}}\sum_{\substack{rstu \\ vwxy}}\psi_{i^2b^2e^2}\psi_{i^3b^3e^3}^* 
 U_{(rs)(tu)}U^*_{(xy)(vw)}\ket{rs}\langle{tu}\ket{i^2b^2e^2}\langle{i^3b^3e^3}\ket{vw}\langle{xy}\ket{(\gamma'\beta')\xi'} \nonumber \\ 
 & = & \sum_{\gamma\gamma'\beta\beta'\xi\xi'}\sum_{\substack{beb'e' \\ b^2e^2b^3e^3}}\psi_{\gamma be}\psi^*_{\gamma' b'e'}\psi_{\gamma'b^2e^2}\psi^*_{\gamma b^3e^3}  \int_H\hbox{dU}\left[ U^*_{(\beta'\xi)(b'e')}U^*_{(\beta\xi')(b^3e^3)}U_{(\beta\xi)(be)}U_{(\beta'\xi')(b^2e^2)}\right] \nonumber \\
 & \equiv & I+J\, ,
\end{eqnarray}
where we have split the sum into two components
\begin{eqnarray}
 I & := & \sum_{\gamma\beta\beta'\xi\xi'}\sum_{n_0n_1n_2n_3}\psi_{\gamma n_0}\psi^*_{\gamma n_1}\psi_{\gamma n_2}\psi^*_{\gamma n_3}\int_H\hbox{dU}\left[ U^*_{(\beta'\xi)n_1}U^*_{(\beta\xi')n_3}U_{(\beta\xi)n_0}U_{(\beta'\xi')n_2}\right]\, , \nonumber \\
 J & := & \sum_{\substack{\gamma\gamma'\beta\beta'\xi\xi' \\ (\gamma\neq\gamma')}}\sum_{n_0n_1n_2n_3}\psi_{\gamma n_0}\psi^*_{\gamma' n_1}\psi_{\gamma' n_2}\psi^*_{\gamma n_3}\int_H\hbox{dU}\left[ U^*_{(\beta'\xi)n_1}U^*_{(\beta\xi')n_3}U_{(\beta\xi)n_0}U_{(\beta'\xi')n_2}\right]\, . \nonumber \\ 
\end{eqnarray}
In the above equalities we have also shortened the notation, merging the boundary and environment indices pertaining to both the coefficients of $\ket{\Psi}$ and $U$ into a single one. 
Let us take care of $I$ first.
\begin{eqnarray} \label{eq:decompose}
 I & = & \underbrace{\sum_{\substack{\gamma\beta\xi\xi' \\ \xi\neq\xi'}}\sum_{n_0n_1n_2n_3}\psi_{\gamma n_0}\psi^*_{\gamma n_1}\psi_{\gamma n_2}\psi^*_{\gamma n_3}\int_H\hbox{dU}\left[ U^*_{(\beta\xi)n_1}U^*_{(\beta\xi')n_3}U_{(\beta\xi)n_0}U_{(\beta\xi')n_2}\right]}_{=:\hbox{\textbf{I}}(\xi\neq\xi')}+ \nonumber\\
   &   & \underbrace{\sum_{\substack{\gamma\beta\beta'\xi \\ \beta\neq\beta'}}\sum_{n_0n_1n_2n_3}\psi_{\gamma n_0}\psi^*_{\gamma n_1}\psi_{\gamma n_2}\psi^*_{\gamma n_3}\int_H\hbox{dU}\left[ U^*_{(\beta'\xi)n_1}U^*_{(\beta\xi)n_3}U_{(\beta\xi)n_0}U_{(\beta'\xi)n_2}\right]}_{=:\hbox{\textbf{II}}(\beta\neq\beta')}+ \nonumber \\
   &   & \underbrace{\sum_{\gamma\beta\xi}\sum_{n_0n_1n_2n_3}\psi_{\gamma n_0}\psi^*_{\gamma n_1}\psi_{\gamma n_2}\psi^*_{\gamma n_3}\int_H\hbox{dU}\left[ U^*_{(\beta\xi)n_1}U^*_{(\beta\xi)n_3}U_{(\beta\xi)n_0}U_{(\beta\xi)n_2}\right]}_{=:\hbox{\textbf{III}}(\beta\xi=\beta'\xi')}\, . \nonumber \\
\end{eqnarray}
We should now further decompose the sums above to get to a sum of integrals like Eq.s(\ref{eq:nonzero1}-\ref{eq:nonzero5}) of Appendix \ref{app:measure}. Writing it down explicitly would be rather unmanageable, though.
Arguably the best way to work it out is to group the decomposition into a table. 
\begin{table}[tb] 
\begin{tabular}{l|c|c|r}
  $\sum_{n_0n_1n_2n_3}$ & \textbf{I}$(\xi\neq\xi')$ & \textbf{II}$(\beta\neq\beta')$ & \textbf{III}$(\beta\xi=\beta'\xi')$ \\
 \hline 
 $n_0=n_1\neq n_2=n_3$ & $d_Bd_E(d_E-1)\cdot $(\ref{eq:nonzero1}) & $d_Bd_E(d_B-1)\cdot $(\ref{eq:nonzero2}) & $d_Bd_E\cdot $(\ref{eq:nonzero3}) \\
 $n_0=n_3\neq n_1=n_2$ & $d_Bd_E(d_E-1)\cdot $(\ref{eq:nonzero2}) & $d_Bd_E(d_B-1)\cdot $(\ref{eq:nonzero1}) & $d_Bd_E\cdot $(\ref{eq:nonzero3}) \\
 $n_0=n_1=n_2=n_3$ & $d_Bd_E(d_E-1)\cdot $(\ref{eq:nonzero4}) & $d_Bd_E(d_B-1)\cdot $(\ref{eq:nonzero4}) & $d_Bd_E\cdot $(\ref{eq:nonzero5}) \\
 \hline 
\end{tabular}
\caption{Decomposition of $I$ into a sum of the non-zero integrals Eqs.(\ref{eq:nonzero1}-\ref{eq:nonzero5}). In the second to fourth columns, each 
         of these factors is multiplied by their multiplicity.} \label{tab:intI}
\end{table}
Using Tab.(\ref{tab:intI}), after a little bookkeeping we obtain
\begin{small}
\begin{eqnarray} \label{eq:Ifinal}
I & = & \left(d(d_E-1)+d(d_B-1)\right)\left( \sum_{\substack{n_0n_2\gamma \\ n_0\neq n_2}}\frac{|\psi_{\gamma n_0}|^2|\psi_{\gamma n_2}|^2}{(d^2-1)} + \sum_{n_0\gamma}\frac{|\psi_{\gamma n_0}|^2|\psi_{\gamma n_0}|^2}{d(d+1)} -  \sum_{\substack{n_0n_2\gamma \\ n_0\neq n_2}}\frac{|\psi_{\gamma n_0}|^2|\psi_{\gamma n_2}|^2}{d(d^2-1)}\right) \nonumber \\
  & + &  2d\left( \sum_{\substack{n_0n_2\gamma \\ n_0\neq n_2}}\frac{|\psi_{\gamma n_0}|^2|\psi_{\gamma n_2}|^2}{d(d+1)}+\sum_{n_0\gamma}\frac{|\psi_{\gamma n_0}|^2|\psi_{\gamma n_0}|^2}{d(d+1)}\right)\,
\end{eqnarray}
\end{small}
with $d=d_Ed_B$.
\\

We can do exactly the same thing with $J$. First, decompose the sum as in Eq.(\ref{eq:decompose})
\begin{eqnarray}
 J & = & \underbrace{\sum_{\substack{\gamma\gamma' \\ \gamma\neq\gamma'}}\sum_{\substack{\beta\xi\xi' \\ \xi\neq\xi'}}\sum_{n_0n_1n_2n_3}\psi_{\gamma n_0}\psi^*_{\gamma' n_1}\psi_{\gamma' n_2}\psi^*_{\gamma n_3}\int_H\hbox{dU}\left[ U^*_{(\beta\xi)n_1}U^*_{(\beta\xi')n_3}U_{(\beta\xi)n_0}U_{(\beta\xi')n_2}\right]}_{\hbox{\textbf{I}}(\xi\neq\xi')}+ \nonumber \\
   &   & \underbrace{\sum_{\substack{\gamma\gamma' \\ \gamma\neq\gamma'}}\sum_{\substack{\beta\beta'\xi \\ \beta\neq\beta'}}\sum_{n_0n_1n_2n_3}\psi_{\gamma n_0}\psi^*_{\gamma' n_1}\psi_{\gamma' n_2}\psi^*_{\gamma n_3}\int_H\hbox{dU}\left[ U^*_{(\beta'\xi)n_1}U^*_{(\beta\xi)n_3}U_{(\beta\xi)n_0}U_{(\beta'\xi)n_2}\right]}_{\hbox{\textbf{II}}(\beta\neq\beta')}+ \nonumber \\
   &   & \underbrace{\sum_{\substack{\gamma\gamma' \\ \gamma\neq\gamma'}}\sum_{\beta\xi}\sum_{n_0n_1n_2n_3}\psi_{\gamma n_0}\psi^*_{\gamma' n_1}\psi_{\gamma' n_2}\psi^*_{\gamma n_3}\int_H\hbox{dU}\left[ U^*_{(\beta\xi)n_1}U^*_{(\beta\xi)n_3}U_{(\beta\xi)n_0}U_{(\beta\xi)n_2}\right]}_{\hbox{\textbf{III}}(\beta\xi=\beta'\xi')}\, , \nonumber \\
\end{eqnarray}
and then use Tab.(\ref{tab:intI}) [the integrals are the same of Eq.(\ref{eq:decompose})] to obtain, after some algebra,
\begin{small}
\begin{eqnarray} \label{eq:Jfinal}
J & = & d(d_E-1)\left( \sum_{\substack{\gamma\gamma' \\ \gamma\neq\gamma'}}\sum_{\substack{n_0n_2 \\ n_0\neq n_2}}\frac{\psi_{\gamma n_0}\psi_{\gamma' n_0}^*\psi_{\gamma' n_2}\psi_{\gamma n_2}^*}{(d^2-1)} - \sum_{\substack{\gamma\gamma' \\ \gamma\neq\gamma'}}\sum_{\substack{n_0n_2 \\ n_0\neq n_2}}\frac{|\psi_{\gamma n_0}|^2|\psi_{\gamma' n_2}|^2}{d(d^2-1)} + \sum_{\substack{n_0 \gamma\gamma' \\ \gamma\neq\gamma'}}\frac{|\psi_{\gamma n_0}|^2|\psi_{\gamma' n_0}|^2}{d(d+1)}\right) \nonumber \\
  & + &  d(d_B-1)\left(\sum_{\substack{\gamma\gamma' \\ \gamma\neq\gamma'}}\sum_{\substack{n_0n_2 \\ n_0\neq n_2}}\frac{|\psi_{\gamma n_0}|^2|\psi_{\gamma' n_2}|^2}{(d^2-1)} - \sum_{\substack{\gamma\gamma' \\ \gamma\neq\gamma'}}\sum_{\substack{n_0n_2 \\ n_0\neq n_2}}\frac{\psi_{\gamma n_0}\psi_{\gamma' n_0}^*\psi_{\gamma' n_2}\psi_{\gamma n_2}^*}{d(d^2-1)}  + \sum_{\substack{n_0 \gamma\gamma' \\ \gamma\neq\gamma'}}\frac{|\psi_{\gamma n_0}|^2|\psi_{\gamma' n_0}|^2}{d(d+1)}\right) \nonumber \\
  & + & \frac{1}{(d+1)} \left( \sum_{\substack{\gamma\gamma' \\ \gamma\neq\gamma'}}\sum_{\substack{n_0n_2 \\ n_0\neq n_2}}\psi_{\gamma n_0}\psi_{\gamma' n_0}^*\psi_{\gamma' n_2}\psi_{\gamma n_2}^* \sum_{\substack{\gamma\gamma' \\ \gamma\neq\gamma'}}\sum_{\substack{n_0n_2 \\ n_0\neq n_2}}|\psi_{\gamma n_0}|^2|\psi_{\gamma' n_2}|^2 +2 \sum_{\substack{n_0 \gamma\gamma' \\ \gamma\neq\gamma'}}|\psi_{\gamma n_0}|^2|\psi_{\gamma' n_0}|^2 \right) .
\end{eqnarray}
\end{small}
Putting it all together and rearranging a bit:
\begin{small}
\begin{eqnarray}
I+J & = & \frac{d_E+d_B}{d+1} \left( \sum_{\substack{\gamma n_0n_2 \\ n_0\neq n_2}}|\psi_{\gamma n_0}|^2|\psi_{\gamma n_2}|^2+\sum_{\gamma n_0}|\psi_{\gamma n_0}|^2|\psi_{\gamma n_0}|^2+ \sum_{\substack{\gamma\gamma' n_0 \\ \gamma\neq\gamma'}}|\psi_{\gamma n_0}|^2|\psi_{\gamma' n_0}|^2 \right) \nonumber \\
 & = & \left[ \frac{dd_B-d_E}{(d-1)(d+1)}\right] \sum_{\substack{\gamma\gamma' \\ \gamma\neq\gamma'}}\sum_{\substack{n_0n_2 \\ n_0\neq n_2}} |\psi_{\gamma n_0}|^2|\psi_{\gamma' n_2}|^2 + \left[ \frac{dd_E-d_B}{(d-1)(d+1)}\right] \sum_{\substack{\gamma\gamma' \\ \gamma\neq\gamma'}}\sum_{\substack{n_0n_2 \\ n_0\neq n_2}} \psi_{\gamma n_0}\psi_{\gamma' n_0}^*\psi_{\gamma' n_2}\psi_{\gamma n_2}^* .
\end{eqnarray}
\end{small}

The calculation to get to the final expression of the average purity is rather tedious, but quite trivial. It only involves some algebra and some care in grouping the right terms to form the trace of the global state. Ultimately, this leads to
\begin{equation}\label{avpur}
 \langle \mu(\rho_{IB})\rangle = \frac{d_B+d_E}{d_Ed_B+1}+\frac{d_B(1-d_E^2)}{(d_Ed_B)^2-1}
 \sum_{\substack{\gamma\gamma' \\ \gamma\neq\gamma'}}\sum_{\substack{n_0n_2 \\ n_0\neq n_2}}\left( |\psi_{\gamma n_0}|^2|\psi_{\gamma' n_2}|^2-\psi_{\gamma n_0}\psi_{\gamma' n_0}^*\psi_{\gamma' n_2}\psi_{\gamma n_2}^* \right)\, .
\end{equation}
\begin{flushright}
 $\square$
\end{flushright}

\section{Mean purity of initially uncorrelated pure states}\label{app:schmidt}

{\em Proof:}
The initial state $\rho_{IBE}=\ketbra{\Psi}{\Psi}$ is such that 
\begin{equation} \label{eq:sep}
\ket{\Psi}=\ket{\Psi_{IB}}\otimes \ket{\Psi_E}=\left( \sum_i^M c_i\ket{I_i}\otimes \ket{B_i}\right)\otimes \left( \sum_j\xi_j\ket{E_j}\right) \, ,
\end{equation}
where $M=\hbox{min}(\left( d_B, d_I \right)$.
By plugging the coefficients of the above equation into Eq.(\ref{eq:purityres}), 
we have that the last term in the sum must ($\Gamma$) be null (this is due to the fact that $\psi_{i(be)}\propto \delta_{ib}$, because of the Schmidt decomposition). 
So we have:
\begin{eqnarray} 
\Delta - \Gamma & = & \sum_{\substack{i,i',(be),(be)'\\ i\neq i',(be)\neq (be)'}}\left[|\psi_{i(be)}|^2|\psi_{i'(be)'}|^2-\psi_{i(be)}\psi_{i'(be)}^*\psi_{i'(be)'}\psi_{i(be)'}^*\right] 
 =  \sum_{\substack{jklm\\ j\neq l}}|c_j\xi_k|^2|c_l\xi_m|^2 \nonumber \\
& = & \sum_{\substack{jl\\ j\neq l}}|c_j|^2|c_l|^2 \sum_k|\xi_k|^2 \sum_m|\xi_m|^2 
 =  \sum_{\substack{jl\\ j\neq l}}|c_j|^2|c_l|^2 \, ,
\end{eqnarray}
where the last equality follows from the fact that the reduced state of the environment is trace-one.
Hence when the initial state is separable such as is Eq.(\ref{eq:sep}), the resulting purity is:
\begin{equation}
\langle \mu(\rho_{IB})\rangle=\frac{d_B+d_E}{d_Bd_E+1}+\frac{d_B(1-d_E^2)}{(d_Bd_E)^2-1}\sum_{\substack{jl\\ j\neq l}}|c_j|^2|c_l|^2 \, . 
\end{equation} 
\begin{flushright}
 $\square$
\end{flushright}

\section{Measure concentration and Levy's lemma} \label{app:levy}

$d$-dimensional pure quantum states can be described as points on the surface of a $(2d-1)$-dimensional unit sphere.
This can be realized by expressing a generic state in complex coordinates $\ket{\psi}=(z_1,z_2,\dots,z_d)$,
where $z_j\in \mathbb{C}$, for $j=1,2,\dots,d$, with $\sum_j^d|z_j|^2=1$ and writing the coordinates in real
components $z_j=x_j+iy_j$, so that $\sum_j^dx_j^2+\sum_j^dy_j^2=1$.

Heuristically, the phenomenon of measure concentration on a unit sphere $S^{(2d-1)}$ in $\mathbb{R}^{2d}$ 
translates to the fact that almost all surface measure of the sphere is concentrated around the equator, for \textit{any}
equator. That is, for any random choice of a coordinate $x_j$, consider an equator of width $\epsilon$
\begin{equation}
 E_{\epsilon}:=\{ x_j\in S^{(2d-1)}\, | \, d(x_j,0)\leq \frac{\epsilon}{2}\}\, ,
\end{equation}
where $d(x,y)=\hbox{arccos}\langle x,y\rangle$ $\forall x,y\in S^{(2d-1)}$ is the angular distance.
Provided a normalised surface measure $\xi(S^{(2d-1)})=1$, it can be shown that 
\begin{equation}
 \xi(E_{\epsilon})\geq 1-e^{(-kd\epsilon^2)}\, ,
\end{equation}
where $k>0$ is some constant.

Measure concentration is at the basis of Levy's lemma (Sec.(\ref{sec:purity})), as we show in the following.
The kind of Levy's lemma we will sketch the derivation of here is slightly different from the one applied in the next section but the two formulations are strictly related. For space reasons,
not all the details of the calculations will be shown; the interested reader can find them in \cite{ledoux}.

To proceed, we first need to define two quantities:
\begin{defn}
 (Median): Let $X$ be a metric space and $f\, :\, X \, \rightarrow \, \mathbb{R}$ a continuous function.
 A median $M_f$ is defined by:
 \begin{equation}
  \xi\{x\in X \, | \, f(x)\leq M_f\}=\frac{1}{2}\, .
 \end{equation}
\end{defn}
\begin{defn}
 (Concentration function): Let $X$ be a metric space and $S$ a subset of it, with $\xi(S)=\frac{1}{2}$.
 For any $\epsilon>0$, the concentration function is defined as:
 \begin{equation}
  \alpha_X(\epsilon):=sup\{\xi(X\setminus N_{\epsilon}(S))\}\, ,            
 \end{equation}
 where $N_{\epsilon}(S)$ is the $\epsilon$-neighborhood of $S$:
 \begin{equation}
  N_{\epsilon}(S):=\{ x\in X\, | \, \exists s\in S\, : \, d(s,x)<\epsilon\}\, .
 \end{equation}
\end{defn}
These definitions allow one to formulate the following lemma:
\begin{lemma} \label{lemma:merda}
 Let $X$ be a metric space and $f\, :\, X \, \rightarrow \, \mathbb{R}$ a Lipschitz-continuous function
 with constant 1, then
 \begin{equation}
  \xi\{x\in X\, | \, f(x)\geq M_f+\epsilon\}\leq \alpha_X(\epsilon) \; .
 \end{equation}
\end{lemma}

{\em Proof:}
Take $S\, :\, \{x\, |\, f(x)\leq M_f\}$ so that $\xi(S)=1/2$ and consider a subset $B\subseteq X$ such that
$f(b)\geq M_f+\epsilon$, $\forall b\in B$. Because $f$ is Lipschitz continuous, all points $x\in N_{\epsilon}(S)$
satisfy $f(x)<M_f+\epsilon$, so
it must be $b\notin N_{\epsilon}(S)$, $\forall b\in B$.
That means $B$ is a subset of $X$: $\{b\in X\, | \, f(b)\geq M_f+\epsilon\}\subseteq X\setminus N_{\epsilon}(S)$ and thus 
$\xi\{x\in X\, | \, f(x)\geq M_f+\epsilon\}\leq\xi(X\setminus N_{\epsilon}(S))\leq\alpha_X(\epsilon)$. 
\begin{flushright}
$\square$
\end{flushright}

In terms of probabilities, and by rescaling of the $\epsilon$ to $\epsilon\rightarrow\epsilon'=\eta\epsilon$ for
Lipschitz functions such that $|f(x)-f(y)|\leq\eta\|x-y\|\leq \eta\epsilon $, the above lemma reads:
\begin{equation}
 \hbox{Prob}(f(x)\geq M_f+\epsilon')\leq\alpha_X\left(\frac{\epsilon'}{\eta}\right)\, .
\end{equation}

In order to calculate the value of the concentration function $\alpha_{S^{(2d-1)}}$, one needs to invoke the isoperimetric inequality
for the sphere (see e.g. \cite{ledoux}):
\begin{lemma}
 (Isoperimetric inequality for the sphere): Let $A\subseteq S^{(2d-1)}$ be a closed subset of the sphere and let 
 $C(a,r):=\{ x\in X \, | \, d(a,x)\leq r\}\subset S^{(2d-1)}$ a spherical cap around any point $a\in S^{(2d-1)}$,
 with the radius r chosen such that $\xi(C(a,r))=\xi(A)$. Then
 \begin{equation}
  \xi(N_{\epsilon}(A))\geq \xi(N_{\epsilon}(C(a,r)))\, .
 \end{equation}
\end{lemma}
Therefore we have
\begin{equation}
 \alpha_{S^{(2d-1)}}(\epsilon)  =  sup\{\xi(S^{(2d-1)}\setminus N_{\epsilon}(S))\}  =  \xi (S^{(2d-1)})-\hbox{inf}\{\xi(N_{\epsilon}(S))\}  =  1-\hbox{inf}\{\xi(N_{\epsilon}(S))\}
  =  1-\xi(C(a,\frac{\pi}{2}+\epsilon))  \leq  e^{-d\epsilon^2}\, ,
\end{equation}
where the details of the calculation leading to the last inequality can be found in \cite{gerken}.

So far then, for functions $f$ with Lipschitz constant $\eta\leq 1$:
\begin{equation}
 \xi\{f(x)\geq M_f+\epsilon\}\leq\alpha_{S^{(2d-1)}}(\epsilon)\leq e^{-d\epsilon^2}\, .
\end{equation}
Applying Lemma \ref{lemma:merda} to the function $g(x)=-f(x)$, one gets
$\xi\{g(x)\geq M_f-\epsilon\}\leq\alpha_{S^{(2d-1)}}(\epsilon)$, thus
\begin{equation}
 \xi\{|f(x)-M_f|\geq\epsilon\}\leq 2\alpha_{S^{(2d-1)}}(\epsilon)\, .
\end{equation}

By rescaling $\epsilon\rightarrow\epsilon\eta$ for functions with $\eta\geq 1$ and interpreting the relative measure
above as a probability, we get to
\begin{equation}
 \hbox{Prob}\{|f(x)-M_f|\geq\epsilon\}\leq2\hbox{e}^{-d\frac{\epsilon^2}{\eta^2}}\, .
\end{equation}
Finally, an inequality can be shown which relates median and expectation value of $f$, bringing the missing
factors in the exponential which appear in the version of the Levy's lemma we made use of in Eq.~(\ref{eq:levydef}).

\section{Bound to state fluctuations}\label{app:flucs}

Let us now make use of Levy's lemma 
to bound the fluctuations around the average, 
unconditional state $\langle\rho_{IB}\rangle=\rho_I\otimes \mathbb{I}/d_B$. 
To this aim, we intend to apply the lemma 
to the trace distance between states, which quantifies 
their operational distinguishability \cite{helstrom,holevo},
in particular between the conditional state $\rho_I\otimes\ket{\phi}\bra{\phi}$ under the assumption of separability between the initial inner and boundary systems [see Eq.~(\ref{conditional})],  and the unconditional, 
average state:
$f(\phi)=\|\rho_I\otimes\hbox{Tr}_E(\ket{\phi}\bra{\phi}) -\rho_I\otimes\mathbb{I}/d_B \|_1=\|\hbox{Tr}_E(\ket{\phi}\bra{\phi}) -\mathbb{I}/d_B \|_1$.   
Levy's lemma applied to $f(\phi)$ reads:
\be
\label{eq:levystraight}
\hbox{P}\big[|\|\hbox{Tr}_E(\ket{\phi}\bra{\phi}) -\mathbb{I}/d_B\|_1 -
\langle\|\hbox{Tr}_E(\ket{\phi}\bra{\phi}) -\mathbb{I}/d_B\|_1\rangle 
\bigr\rvert\geq\epsilon\big]
\leq 2\hbox{exp}\left(-\frac{2d_Bd_E\epsilon^2}{9\pi^3\eta^2}\right)\, .
\ee

It is convenient to rearrange Eq.(\ref{eq:levystraight}) such that we get to an expression of the form:
\begin{equation} \label{eq:levyrearranged}
\hbox{P}\left[ \|\hbox{Tr}_E(\ket{\phi}\bra{\phi}) -\mathbb{I}/d_B \|_1 \geq\gamma\right]\leq \gamma'\, ,
\end{equation}
where
\begin{equation}
\gamma=\epsilon + \langle\|\hbox{Tr}_E(\ket{\phi}\bra{\phi}) -\mathbb{I}/d_B \|_1\rangle\, , \quad
\gamma'= 2\hbox{exp}\left(-\frac{2d_Bd_E\epsilon^2}{9\pi^3\eta^2}\right)\, .
\end{equation}
So in order to estimate the fluctuations around the mean state we need to bound $\langle\|\hbox{Tr}_E(\ket{\phi}\bra{\phi}) -\mathbb{I}/d_B \|_1\rangle$.
Following an argument presented in \cite{popescuNature},
it is now convenient to turn to the more accessible Hilbert-Schmidt norm $\|M\|_2=\sqrt{\hbox{Tr}(M^{\dagger}M)}$ by exploiting its relationship with the trace norm 
$\|M\|_1=\hbox{Tr}\sqrt{M^{\dagger}M}$, which satisfies, for any $n\times n$ matrix $M$, the relation $\|M\|_1^2\leq n\|M\|_2^2$.
We thus have:
\begin{eqnarray}
\langle\|\hbox{Tr}_E(\ket{\phi}\bra{\phi}) -\mathbb{I}/d_B \|_1\rangle &\le& 
\sqrt{d_B}\langle\|\hbox{Tr}_E(\ket{\phi}\bra{\phi}) -\mathbb{I}/d_B \|_2\rangle \le
\sqrt{d_B \langle\|\hbox{Tr}_E(\ket{\phi}\bra{\phi}) -\mathbb{I}/d_B \|_2^2\rangle} \nonumber \\
&=& \sqrt{d_B\langle\hbox{Tr}\left[\left(\hbox{Tr}_E(\ket{\phi}\bra{\phi}) -\mathbb{I}/d_B\right)^2\right]\rangle}
= \sqrt{d_B \langle \hbox{Tr}_B\left[\left(\hbox{Tr}_E(\ket{\phi}\bra{\phi})\right)^2\right]\rangle - 1}\\ &=&
\sqrt{\frac{d_{B}^2-1}{d_Bd_E+1}}\, , \nonumber
\end{eqnarray}
where we inserted the average local purity $(d_B+d_E)/(d_Bd_E+1)$ in the absence of initial $IB$ correlations [see Eq.~(\ref{avpur})].

To complete the Levy's bound we are looking for, we make use of the following lemma \cite{popescuNature}:
\begin{lemma} \label{lemma:lipmap}
The Lipschitz constant $\eta$ of the function 
$f(\phi)=\|\hbox{Tr}_E\left[\ket{\phi}\bra{\phi}-\right]-\mathbb{I}/d_B\|_1$
satisfies $\eta\leq 2$.
\end{lemma}
{\em Proof:} One has 
\begin{align}
|f(\phi_1)-f(\phi_2)|^2 & = 
|\|\hbox{Tr}_E\left[\ket{\phi_1}\bra{\phi_1}\right]-\mathbb{I}/d_B\|_1 -
\|\hbox{Tr}_E\left[\ket{\phi_2}\bra{\phi_2}\right]-\mathbb{I}/d_B\|_1|^2 \le 
\|\hbox{Tr}_E\left[\ket{\phi_1}\bra{\phi_1}-\ket{\phi_2}\bra{\phi_2}\right]\|_1^2 \nonumber \\
\le & \|\ket{\phi_1}\bra{\phi_1}-\ket{\phi_2}\bra{\phi_2}\|_1
= 4\left(1- |\bra{\phi_1}\phi_2\rangle|^2\right)
\le 4|\ket{\phi_1}-\ket{\phi_2}|^2 \; ,
\end{align}
where we used the non-increase of the trace norm under partial tracing as well as the reverse triangle inequality. The last inequality is equivalent to $\eta\le 2$.
\begin{flushright}
 $\square$ 
\end{flushright}

The lemma above allows one to upper bound the quantity $\gamma'$ in Eq.(\ref{eq:levyrearranged}) and thus obtain Inequality (\ref{eq:lastlevy})  (where $\rho_{IB,\phi}\equiv\rho_I\otimes \hbox{Tr}[\ket{\phi}\bra{\phi}]$):
\begin{equation} 
\hbox{P}\left[ \|\rho_I\otimes \hbox{Tr}[\ket{\phi}\bra{\phi}] -\Phi \left( \rho_{\phi}\right) \|_1 \geq \epsilon + \sqrt{\frac{d_B^2-1}{d_Bd_E+1}}\right] \leq 2\hbox{exp}\left(-\frac{d_Bd_E\epsilon^2}{18\pi^3}\right) \, , 
\end{equation}
which bounds the fluctuations around the average state of 
Eq.~(\ref{eq:propmap}).

\end{document}